\documentclass[9pt,conference]{IEEEtran}

\usepackage[preprint]{waspaa25}

\usepackage{bm} %

\newif\iftodo

\title{Combolutional Neural Networks}

\name{Cameron Churchwell,
      Minje Kim,
      Paris Smaragdis}
\address{University of Illinois at Urbana-Champaign, Siebel School of Computing and Data Science, Urbana IL, USA, 61801}

\begin{document}

\maketitle

\begin{abstract}

Selecting appropriate inductive biases is an essential step in the design of machine learning models, especially when working with audio, where even short clips may contain millions of samples. To this end, we propose the combolutional layer: a learned-delay IIR comb filter and fused envelope detector, which extracts harmonic features in the time domain. We demonstrate the efficacy of the combolutional layer on three information retrieval tasks, evaluate its computational cost relative to other audio frontends, and provide efficient implementations for training. We find that the combolutional layer is an effective replacement for convolutional layers in audio tasks where precise harmonic analysis is important, e.g., piano transcription, speaker classification, and key detection. Additionally, the combolutional layer has several other key benefits over existing frontends, namely: low parameter count, efficient CPU inference, strictly real-valued computations, and improved interpretability.
\end{abstract}

\section{Introduction}

Feature engineering in machine learning has largely been replaced by model architectures with inductive biases which encode knowledge of trends in the target distribution. For example, convolutional networks classify handwritten digits with equal accuracy to fully-connected networks while requiring significantly fewer parameters, which also reduces training complexity~\cite{leGOAT}. By selecting an appropriate inductive bias, we bypass the need to hand-craft features and allow the model to learn more tailored features as part of the model training: CNNs encode dependencies of adjacent samples \cite{CNN}; RNNs encode backward dependencies in time~\cite{RNN}; LSTMs encode longer-term backward dependencies~\cite{LSTM}; Transformers encode dependencies across entire sequences~\cite{Attention}. However, the improvements in efficiency enabled by these models do not erase the challenges inherent in modeling long sequences of audio samples.

Many models avoid these challenges by relying on dimensionality reduction through time-frequency domain transforms and hand-crafted features to attain state-of-the-art performance~\cite{whisper,AST,vocos}. The mel spectrogram is an especially popular choice for speech-related tasks as it utilizes frequency scaling informed by human auditory perception\cite{melgan,hifigan}. While this is an effective approach for many tasks, the exact choice of features used can lead to loss of task-critical information, which, in turn, can have significant impacts on model performance~\cite{features}.

To avoid task-feature mismatch issues with hand-crafted features, a more recent approach is to learn features from the raw inputs. Models such as Conv-TasNet for source separation~\cite{convtasnet}, Wav2vec for self-supervised feature learning~\cite{wav2vec}, and WaveNet for neural vocoding~\cite{wavenet}, operate on waveforms directly using convolutional layers, capitalizing on the relatedness of neighboring samples to learn relationships which might not be captured by DSP-style features. The choice of CNNs to model audio is strongly supported by signal processing theory, as convolutional layers are equivalent to learned finite impulse response (FIR) filters \cite{DDSP}. Notably, the filters learned by these models are unconstrained apart from by their kernel size, in the same way that a fully-connected network is unconstrained compared to a CNN. Meanwhile, convolutional layers are not always the most suitable operation for signal processing, often requiring a large model to properly handle the learning task involving raw waveform signals \cite{wavenet}.

The practice of parameterizing DSP elements for use in neural networks is commonly referred to as differentiable digital signal processing (DDSP)~\cite{DDSP}. The most well known DDSP application is to use neural networks to predict synthesizer parameters to reconstruct realistic audio of musical instruments \cite{MIDIDDSP}. Other works have demonstrated the potential for machine learning models to learn DSP filters, either to digitalize existing analog filters~\cite{DiffIIR,DiffAllPole} or to learn input features for a downstream task, such as SincNet~\cite{sincnet}, where bandpass filters with parameterized cutoff frequencies are incorporated as part of the neural network. This acts as an additional degree of weight-sharing which further reduces parameter count and improves interpretability. Based on this line of work, we propose a simple and efficient neural network layer which parameterizes comb filters, and use it for applications where an efficient analysis of periodic sound components is crucial. 

In this work we propose the use of feedback comb filters for audio modeling applications as a replacement for unconstrained convolutional layers. The feedback (IIR) comb filter passes the original signal through and sums it with a delayed and scaled copy of its own output. For a delay of $K$ samples and a feedback gain of $0<\alpha<1$, the filter is written as:
\begin{align}
    y[n]&=x[n]+\alpha y[n-K] \label{eq:iir}
\end{align}

Combined with an envelope detector, the feedback comb filter may be used as an efficient, learned harmonic feature extractor we call the \textit{combolutional layer}. The combolutional layer offers several key benefits over other audio model frontends: each output channel of the combolutional layer requires only a single parameter; each output sample requires only a single multiply-accumulate (MAC) operation at inference time, making combolutional layers especially attractive for on-device applications.

Our primary contributions are as follows:
\begin{itemize}
    \item We introduce the combolutional layer, a novel parameteric, thus trainable, function that replicates the behavior of a comb filter. Along with other neural network layers, a combolutional network (CombNet) that employs a combolutional layer works as a learnable harmonic feature extractor.
    \item We show that the combolutional layer is more efficient both in terms of parameter count and MACs than an equivalent convolutional layer.
    \item We provide efficient Triton implementations of the non-recursive proxy of the combolutional layer to be used during training. In this way, a combolutional layer can also leverage GPU computing.
    \item We empirically show that CombNets compete with other ConvNet architectures and more efficient ones, such as SincNet, on three harmonically sensitive audio tasks, showcasing efficient inference-time properties. 
\end{itemize}

\section{Comb Filters as Feature Extractors}
\label{sec:feature_extraction}

\begin{figure}[t]
  \centering
  \centerline{\includegraphics[trim=7 5 2 1, clip, width=0.95\columnwidth]{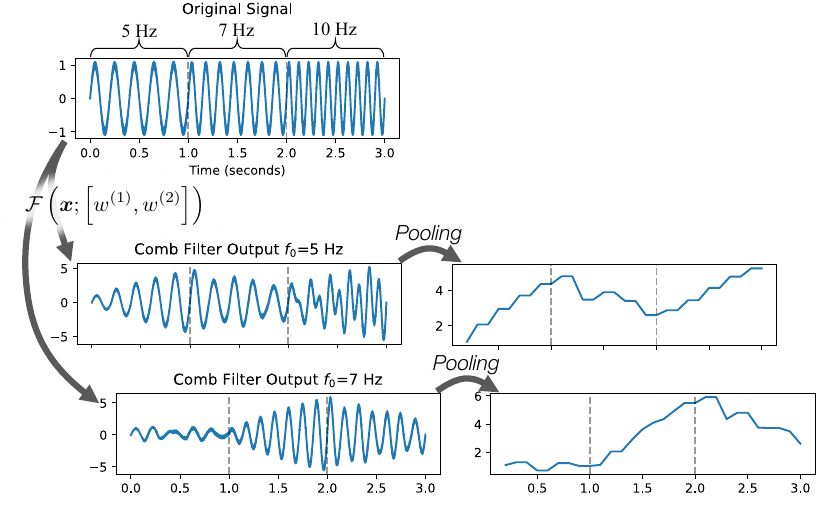}}
  \caption{A combolutional layer with two internal comb filters applied to an input signal comprised of three sequential tones of equal duration at \SIlist{5;7;10}{\hertz}.}
  \label{fig:combolutional}
\end{figure}

\subsection{Definition of Comb Filters}
\label{sec:comb_filter_definition}

The magnitude response of the feedback (IIR) comb filter with a gain of $0<\alpha<1$, sampling rate $f_s$, and delay of $K$ samples, is given by:
\begin{align}
    |H(f)|=\frac{1}{\sqrt{1+\alpha^2-2\alpha\cos(2\pi f/f_s K)}}
\end{align}

If we choose $K=f_s/f_0$, we can rewrite this in terms of a tunable frequency parameter $f_0$:
\begin{align}
    |H(f)|=\frac{1}{\sqrt{1+\alpha^2-2\alpha\cos(2\pi f/f_0)}}
\end{align}

In simple terms, this filter amplifies signal components with frequencies near to integer multiples of $f_0$ while attenuating all other frequencies. Our goal in developing combolutional layers is to replace the comb filter with a surrogate function which is continuous and differentiable in $f_0$. As a result, we obtain a learnable filter with a single parameter, $f_0$, that amplifies the frequencies in the vicinities of $f_0$ and all of its harmonics. As with most neural network layers, a combolutional layer can learn multiple channels in parallel, each corresponding to a different comb filter with its own $f_0$ value. This learned comb filter bank can be combined with the typical activation functions, max pooling, and other additional operations as necessary. 

A significant benefit of learning such a combolutional layer is that it can learn a small and custom set of filters tailored for the task rather than using an over-complete dictionary of many filters. In addition, compared to a convolutional layer, the inherent periodicity of the filters can introduce an inductive bias as well as an additional efficiency to the learning task, mitigating the effort of learning such a structure within a convolutional layer. It should be noted that these advantages come with the cost of the potentially suboptimal assumption that the analysis of the input signal benefits from the periodic filters. Fig.~\ref{fig:combolutional} depicts the harmonic nature of the combolutional layer feature representation. The filter with $f_0=\SI{5}{\hertz}$ detects activity for both the $\SI{5}{\hertz}$ and $\SI{10}{\hertz}$ sine waves at the beginning and end of the signal, but is not too effective in detecting the $\SI{7}{\hertz}$ sine wave. Note that the time-lag shown in the right panel of \ref{fig:combolutional} is exaggerated due to the relatively low frequencies and correspondingly large filter delays.

\subsection{Definition of Combolutional Layers}
\label{sec:comb_layer}

A combolutional layer can be defined as a parametric function $\mathcal{F}(\cdot)$ that takes input signal $\bm x\in\mathbb{R}^T$ and calculates the filter responses $\bm y$:
\begin{equation}
    \bm Y\leftarrow\mathcal{F}(\bm x; \bm w),
\end{equation}
where $\bm w=[w^{(1)}, w^{(2)}, \ldots, w^{(M)}]\in\mathbb{R}^M$ denotes the set of $M$ trainable model parameters, representing the fundamental frequencies of $M$ combolutional kernels. Consequently, there are $M$ output channels as a result of $M$ simultaneous filtering operations, i.e., $\bm Y\in\mathbb{R}^{M\times T'}$, where $T'$ is the length of the response signal. We then take the absolute value and apply max pooling as shown in Fig.~\ref{fig:combolutional}.

Since fundamental frequencies $f_0$ are defined as nonnegative continuous value in Hz, instead of optimizing the model for $f_0$ directly, we use a simple scaling function $s: \mathbb{R}\rightarrow \left[f_{\text{min}}, f_{\text{max}}\right]$ that converts the freely learned model parameter $w^{(m)}$ into its corresponding fundamental frequency value if necessary:
\begin{equation}\label{eq:f_0_conversion}
    f_0^{(m)} = s\left(w^{(m)}\right) = f_{\text{min}} \cdot \left(\frac{f_{\text{max}}}{f_{\text{min}}}\right)^{\sigma \big(w^{(m)}\big)},
\end{equation}
where $\sigma(\cdot)$ is the sigmoid function to introduce nonnegativity into conversion, and $f_{\text{min}}$ and $f_{\text{max}}$ are pre-defined hyperparameters, i.e., the frequency range that the input signal can potentially cover. Note that the exponential scaling is important for training, as a small change $\Delta w^{(m)}$ will result in a roughly equivalent perceptual change $\Delta f_0^{(m)}$, invariant to the current value of $w^{(m)}$.

\section{Efficient Implementations}

Eq.~\eqref{eq:iir} shows the typical discrete computation of a feedback comb filter. While this implementation may be fine for inference time, the discrete nature is not compatible with automatic differentiation packages, nor is it efficient on GPU hardware, where inner product-based parallelization is crucial for efficiency.

\subsection{Non-Recurrent Approximation}

As the first step toward efficient GPU computation, during training, the comb filters are approximated by FIR surrogates via truncated impulse response estimation. The kernel for these filters is straightforward once $\alpha$ and $K$ are defined:
\begin{align}\label{eq:kernel}
    h_K^{(m)}[n] = \begin{cases}
        \alpha^{t} & \text{ if } n = tK, \quad t\in \mathbb{Z}^+ \\
        0 & \text{ otherwise}.
    \end{cases}
\end{align}
In practice, once a combolutional layer parameter $\bm w^{(m)}$ is updated, it is converted to corresponding $f_0^{(m)}$ using eq. \eqref{eq:f_0_conversion}. Based on the known relationship $K=f_s/f_0$, $K$ is inferred. $\alpha$ is a hyperparameter we set to $0.9$ for all experiments.  

At each forward pass, this kernel can be used in a standard convolution:
\begin{align}
    \bm Y_{m,:} \approx h_K^{(m)} * \bm x.
\end{align}
Now, the computation no longer depends on $y$, and so can be fully parallelized across timesteps, which is critical given the alternative is a recurrent operation with potentially millions of time steps.

\subsection{Differentiable Proxy}

Another important step is to relax the definition of comb filtering eq. \eqref{eq:iir} with a discrete delay $K$ in a continuous version to enable gradient computation. This is because during feedforward, the conversion eq. \eqref{eq:f_0_conversion} results in a continuous $f_0$ value, leading to a continuous delay value, while the modulo operation in eq. \eqref{eq:kernel} is defined with an integer $K$. For clarity, we introduce $\bar{K}$ as a continuous version of $K$ learned during training.

To resolve this issue, we use interpolation between two filter responses based on two adjacent integer delays $\lfloor\bar{K}\rfloor$ and $\lceil\bar{K}\rceil$ to compute a continuous proxy of the comb filter with a continuous delay $\bar K$. We find experimentally that linear interpolation between the two is sufficient to compute approximate gradients, turning the filtering into a differentiable operation. The interpolation is defined as follows:
\begin{align}\label{eq:diff_prox}
    \bm Y_{m,:} \approx \big(1-\beta(\bar{K})\big)~h_{\lfloor \bar{K} \rfloor}^{(m)}*\bm x+\beta(\bar{K})~h_{\lceil \bar{K} \rceil}^{(m)}*\bm x,
\end{align}
where $\beta(\bar{K})=\bar{K}-\lfloor \bar{K} \rfloor$ means the fractional part of $\bar K$, which is complementary to the interpolation ratio.

\subsection{Sparsity}

Since the kernel $h_K$ is sparse, we can avoid computational redundancy by implementing the operation as a sum of scaled slices rather than as a typical convolution. To illustrate this point, the following is the discrete (non-differentiable) filter computed via this approach:
\begin{align}
    y[n]=x[n] + \sum_{t=1}^{T} \alpha^t x[n-tK]. \label{eq:discrete_sparse}
\end{align}

By appying the differentiable proxy (eq. \eqref{eq:diff_prox}) to eq.~\eqref{eq:discrete_sparse}, the full implementation of the comb filters used in training becomes:
\begin{align}
    y[n]=x[n]+\sum_{t=1}^T&\big( 1 - \beta\left(t\bar{K}\right)\big) \alpha^t x [ n - \lfloor t\bar{K} \rfloor ] \notag \\
    &+ \beta\left(t\bar{K}\right) \alpha^t x [ n - \lceil t\bar{K} \rceil ].
\end{align}

We provide efficient Triton \cite{triton} kernels and torch forward/backward definitions for the differentiable comb filter as well as a fused combolution layer which is drastically more memory efficient.
~\footnote{\href{https://github.com/CameronChurchwell/combnet}{https://github.com/CameronChurchwell/combnet}}

\subsection{Inference}

During inference, gradient computation is unnecessary and can be removed. This also means that filter $f_0$ values can be discretized for a small performance penalty, and each output sample can be computed sequentially as in eq.~\eqref{eq:iir}. This is particularly ideal for models which must run in real time on-device, as many of these devices are already optimized for IIR filters.

\section{Experiments}

We evaluate the performance of the comb filters on three tasks: monophonic note transcription,  speaker classification \cite{sincnet}, and acoustic musical key estimation. All models were trained on a single Nvidia L40S GPU using the Adam optimizer \cite{adam} and clipping gradients at a value of 0.5. For all comb filters, we use a fixed $\alpha$ value of $0.9$ and a fixed $T$ value of $10$, i.e., we use 10 echoes in our truncated impulse response as written in eq.~(\ref{eq:discrete_sparse}).

\subsection{Note Transcription}

As an illustrative example, we create a synthesized dataset of monophonic piano note sequences, each with a random number of notes between 3 and 10. The notes are sampled uniformly between C4 and B4, also with random durations and velocities%
\footnote{Further details can be found in \href{https://github.com/CameronChurchwell/combnet/blob/main/combnet/data/synthesize/notes.py}{combnet/data/synthesize/notes.py}}.
 
We train two architectures: ConvNet  is composed of three 1D convolutional layers using ELU activations~\cite{ELU}, while the proposed CombNet replaces ConvNet's first layer with a combolutional layer. We train both of these archtectures with 8, 16, 32, 64, and 128 channels in the first and second layer. The final layer always has 12 output channels, one for each of the 12 notes being estimated. All CombNet models use $f_{\text{min}}=\SI{200}{\hertz}$, $f_{\text{max}}=\SI{500}{\hertz}$ for this experiment.

\subsection{Speaker Classification}

We base our speaker classification experiments closely off of the 462-speaker TIMIT~\cite{timit} classification experiment outlined in \cite{sincnet}. We reimplement their SincNet and ConvNet models as baselines and compare them with a CombNet model, which is identical to SincNet except for the replacement of the first sinc layer with a combolutional layer. All models have a selected first layer, then 2 convolutional layers followed by 3 fully-connected layers and a classifier head. We use the same normalization \cite{layernorm,batchnorm}, pooling, and activation \cite{leaky} layers, but swap out the optimizer for Adam~\cite{adam}. SincNet performs well on this task due to its well-aligned inductive bias which can learn the formant bands critical for speaker classification. We show that CombNet also has a powerful inductive bias and is much more efficient at inference time as can be seen in table~\ref{tab:speaker_classification_results}. CombNet uses $f_{\text{min}}=\SI{50}{\hertz}$, $f_{\text{max}}=\SI{8000}{\hertz}$ for this experiment.

\begin{table}[t]
\centering
\caption{Note transcription results. MACs are computed per sample for the first layer only, as the remainder of the network takes multiple samples and process them as a batch, hence their MACs/sample are negligible.}
\label{tab:note_transcription_results}
\sisetup{
    reset-text-series = false, 
    text-series-to-math = true, 
    mode=text,
    tight-spacing=true,
    round-mode=places,
    round-precision=2,
    table-number-alignment=center,
}
\begin{tabular}{l@{\hskip 0pt}*{1}{S[round-precision=2,table-format=2.2]}*{2}{S[round-precision=0,table-format=2]}}
    \toprule
    & {F1 $\uparrow$}& {MACs/sample $\downarrow$}& {Parameters $\downarrow$}\\
    \midrule
    $\text{CombNet}_{128}$ &            0.946679 &            166.84 &           18K \\
    $\text{CombNet}_{64}$ &            0.941933 &            83.42 &           5K \\
    $\text{CombNet}_{32}$ &            0.933040 &            41.71 &           1K \\
    $\text{CombNet}_{16}$ &            0.905495 &            20.86 &           492 \\
    $\text{CombNet}_{8}$ &            0.652382 &           10.43 &           188 \\
    \midrule
    $\text{ConvNet}_{128}$ &            0.951320 &            7M &           1M \\
    $\text{ConvNet}_{64}$ &            0.9461490 &            3M &           569K \\
    $\text{ConvNet}_{32}$ &            0.921002 &            1M &           283K \\
    $\text{ConvNet}_{16}$ &            0.857245 &            987K &           141K \\
    $\text{ConvNet}_{8}$ &            0.747126 &            493K &           70K \\
    \bottomrule
\end{tabular}
\end{table}

\subsection{Key Estimation}

For musical key estimation, we use the GiantSteps-MTG dataset for training and validation, and the GiantSteps \cite{giantsteps} dataset for testing. This task involves predicting one of 24 major and minor keys given a two-minute excerpt of a song. The baseline, $\text{CK}$~\cite{Korzeniowski} takes a $4097$ bin magnitude spectrogram (i.e., the DFT size is $8192$), filters it with $105$ hand-crafted triangular filters into quartertone bins, and then applies a series of 2D convolutions followed by adaptive pooling and a classifier head. We also train three other baselines adapted from $\text{CK}$:
\begin{enumerate} 
    \item $\text{CK}_{\text{tri}}$: replaces the $105$ triangular filters with a learned $4097\times 105$ filter matrix.
    \item $\text{CK}_{\text{learned}}$: Same as $\text{CK}_{\text{tri}}$, also learns a $4097\times 4097$ replacement for the spectrogram computation.
    \item $\text{CK}_{\text{chroma}}$: uses $12$ chroma filters instead of $105$ quartertone filters.
\end{enumerate}
For our combolutional network, $\text{CombNet}_{64}$, we replace the spectrogram and filters with a single combolutional layer with $64$ channels. Then, $\text{CombNet}_{64}$ mimics the hop size and DFT length of $\text{CK}$ as its max pooling stride and window size, respectively. Given the harmonic-centric nature of the key detection problem, we expect that the $64$ harmonic features learned by $\text{CombNet}$ should perform comparably to the $105$ filters used in $\text{CK}$. We use both accuracy and a weighted sum of prediction categories as proposed by the MIREX evaluation campaign\footnote{http://www.music-ir.org/mirex} and also used in \cite{Korzeniowski}. Combnet uses $f_{\text{min}}=\SI{25.95}{\hertz}$, $f_{\text{max}}=\SI{1046.5}{\hertz}$ to match the range of $\text{CK}$

\section{Results}

\subsection{Note Transcription}

\begin{figure}[t]
    \centering
    \centerline{\includegraphics[trim=7 7 7 7, clip, width=1.\columnwidth]{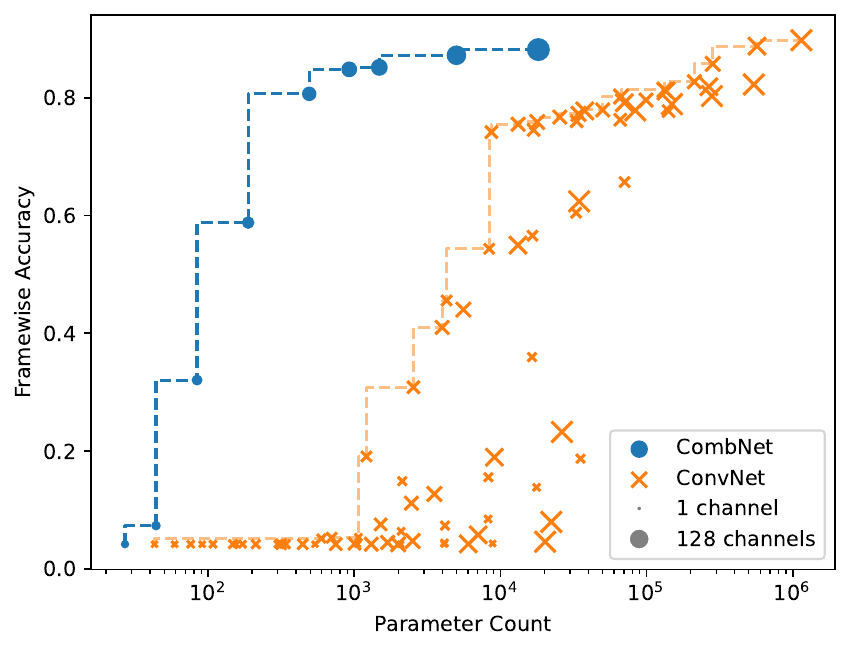}}
    \caption{Pareto fronts for $\text{CombNet}$ and $\text{ConvNet}$ evaluated on the piano transcription task.}
    \label{fig:param_pareto}
  \end{figure}

Table \ref{tab:note_transcription_results} shows the evaluation results for the note transcription task. It is clear from these results that \text{CombNet} is orders of magnitude more efficient than \text{ConvNet} in terms of both parameter count and inference-time multiply-accumulate operations. In particular, $\text{CombNet}_{128}$ and $\text{ConvNet}_{128}$ both achieve an F1 score of $0.95$ but $\text{ConvNet}_{128}$ requires more than four orders of magnitude more MACs and has two orders of magnitude more parameters. In addition, Fig. \ref{fig:param_pareto} shows that across all of the models we trained, \text{CombNet} always achieved superior performance at any given model size.

\begin{figure}[t]
    \centering
    \centerline{\includegraphics[trim=7 7 7 7, clip, width=.9\columnwidth]{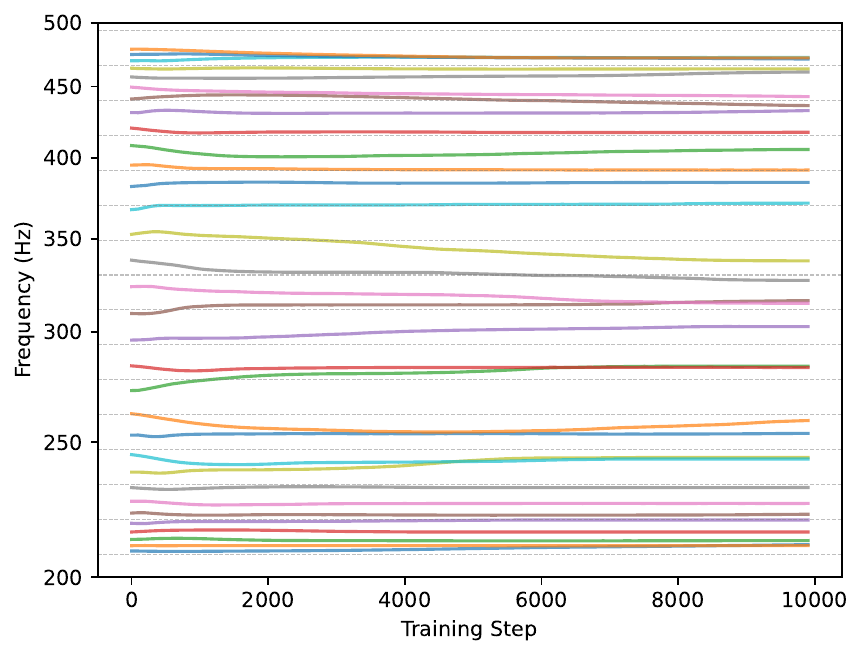}}
    \caption{Evolution of the combolutional layer $f_0$ parameters of $\text{CombNet}_{32}$ during training on the note transcription task.}
    \label{fig:f0_evolution}
  \end{figure}

Fig.~\ref{fig:f0_evolution} shows the process of learning $f_0$ parameters for the combolutional layer when performing note transcription. We observe that multiple filter channels will often converge to the same $f_0$ value, perhaps indicating a particularly useful harmonic feature. 
CombNet also exhibits octave confusion as in other $f_0$ estimation systems. For example, two filters converge to $\sim\SI{243.5}{\hertz}$ whose first harmonic at $\SI{487}{\hertz}$ is close to B4 (\SI{493.88}{\hertz}). Meanwhile, it could be said that CombNet is still efficient when octave confusion is not immediately problematic as in this toy transcription problem.

\subsection{Speaker Classification}

Table \ref{tab:speaker_classification_results} shows the speaker classification evaluation results. $\text{CombNet}_{80}$ and $\text{SincNet}_{80}$ both outperform the $\text{ConvNet}_{80}$ baseline, with $\text{SincNet}_{80}$ performing better than $\text{CombNet}_{80}$ by 2.45 percentage points. However, $\text{SincNet}_{80}$'s slight increase in accuracy comes at the cost of 100 times as many inference MACs for the first layer alone. While each sinc layer only needs two parameters, one for each of the high and low cutoff frequencies of the learned bandpass filter, the computation still involves a convolution with a large FIR filter kernel. The sinc kernels are symmetrical, and so halve the computational cost of an equivalent convolutional layer. However, we claim that CombNet's IIR comb filter is equivalent to a length 1 kernel convolution, and is therefore more efficient. Speaker classification is a task which is frequently performed on-device, and therefore could benefit greatly from this drastic increase in efficiency, even if it comes with a slight performance penalty.

\begin{table}[t]
\centering
\caption{Speaker classification results from CombNet, SincNet, and ConvNet. SincNet and ConvNet results are only slightly different from the originally reported ones in \cite{sincnet} due to reimplementation. MACs are normalized by the number of input audio samples.}
\label{tab:speaker_classification_results}
\sisetup{
    reset-text-series = false, 
    text-series-to-math = true, 
    mode=text,
    tight-spacing=true,
    round-mode=places,
    round-precision=2,
    table-number-alignment=center,
}
\begin{tabular}{l@{\hskip 0pt}*{1}{S[round-precision=2,table-format=2.2]}*{2}{S[round-precision=0,table-format=2]}}
    \toprule
    & Accuracy~\%~$\uparrow$  & \multicolumn{2}{c}{MACs/sample $\downarrow$}\\
    \cmidrule(lr){3-4}
    & & \multicolumn{1}{c}{First Layer} & \multicolumn{1}{c}{Total} \\
    \midrule
    $\text{CombNet}_{80}$ &            96.18 &            80 & 17K \\
    $\text{SincNet}_{80}$ \cite{sincnet} &            98.63 &            9K & 25K \\
    $\text{ConvNet}_{80}$ \cite{sincnet} &            94.52 &            18K & 34K \\
    \bottomrule
\end{tabular}
\end{table}

\begin{table}[t]
\centering
\caption{Key estimation results on the GiantSteps dataset.}
\label{tab:key_estimation_results}
\sisetup{
    reset-text-series = false, 
    text-series-to-math = true, 
    mode=text,
    tight-spacing=true,
    round-mode=places,
    round-precision=2,
    table-number-alignment=center,
}
\begin{tabular}{l@{\hskip 0pt}*{2}{S[round-precision=2,table-format=2.2]}*{1}{S[round-precision=0,table-format=2]}}
    \toprule
    & {Accuracy~\%~$\uparrow$} & {Weighted $\uparrow$} & {Parameters $\downarrow$}\\
    \midrule
    $\text{CombNet}_{64}$ &            61.0927 &           68.8907 &            32K \\
    \midrule
    $\text{CK}$ &            64.4040 &           70.4139 &            48K \\
    $\text{CK}_{\text{tri}}$ &            63.2450 &           69.4040 &            478K \\
    $\text{CK}_{\text{learned}}$ &            40.2318 &           51.1093 &            17M \\
    \midrule
    $\text{CK}_{\text{chroma}}$ &            61.7550 &           68.7583 &            12K \\
    \bottomrule
\end{tabular}
\end{table}

\subsection{Key Estimation}

The results for the key estimation task are shown in Table~\ref{tab:key_estimation_results}. $\text{CK}$, with its hand-crafted filterbanks, performs the best out of all the models trained. However, $\text{CombNet}_{64}$, $\text{CK}_{\text{tri}}$, and $\text{CK}_{\text{chroma}}$ are all within 1.65 points weighted score of $\text{CK}$. We consider this to support the claim that the learned harmonic features of the combolutional layer are equally as useful for machine learning tasks as more traditional DSP-style features. It is also worth noting that the reduction in channel count from $105$ to the $64$ in $\text{CombNet}_{64}$ reduces the parameter count by approximately one quarter, losing only to the highly-customized chroma feature-based $\text{CK}_{\text{chroma}}$ in terms of parameter count.

The relatively strong performance of $\text{CK}_{\text{tri}}$ and comparatively weak performance of $\text{CK}_{\text{learned}}$ indicate that the choice of DSP features is not necessarily as important as the use of a time-frequency transform, as $\text{CK}_{\text{learned}}$ seems to have failed to learn an equally useful representation despite its high parameter count. This reinforces the utility of the combolutional layer as an efficient, trainable time-frequency transform.

\section{Conclusion and Future Work}

In this work we presented combolutional layers as efficient harmonic feature learners. When compared against convolutional frontends, the combolutional layer requires signficantly fewer parameters and is drastically cheaper to compute. We also explore efficient training implementations which enable gradient computation and avoid recurrence, which we open-source. We demonstrated experimentally that combolutional layers perform roughly equivalently to hand-crafted features, and outperformed even very large convolutional layers. We further showed that the combolutional layer can be used on both speech and music, showcasing its broad applicability to harmonic-centric tasks. Our key estimation and speaker classification tasks were minimally different from the baseline sources, which indicates the combolutional layer can be used as a drop-in replacement for other frontends. In terms of future work, we will further investigate the effect of the hyperparameter $\alpha$ and the potential merit of making it trainable.

\clearpage
\bibliographystyle{IEEEtran}
\bibliography{refs}

\end{document}